\documentclass[a4paper,fleqn,usenatbib]{aastex61}

\usepackage{newtxtext,newtxmath}

\usepackage[T1]{fontenc}
\usepackage{ae,aecompl}
 \usepackage[bookmarks=false]{hyperref}

\usepackage{graphicx}	
\usepackage{amsmath}	
\usepackage{amssymb}	
\usepackage{tikz}
\usetikzlibrary{shapes,arrows,positioning,plotmarks,patterns,decorations.pathmorphing,decorations.markings,intersections}
\usepackage{pgf}
\usepackage{pgfplots}
\usepackage{units}

\newcommand\aastex{AAS\TeX}

\received{December 15, 2018}
\submitjournal{PASP}
\usepackage{natbib}
\usepackage{hyperref}
\usepackage{cleveref}

\shorttitle{\aastex\ sample article}
\shortauthors{Rodeghiero et al.}

\begin{document}

\title{Development of the Warm Astrometric Mask for MICADO astrometry calibration}

\correspondingauthor{Gabriele Rodeghiero}
\email{rodeghiero@mpia.de}

\author[0000-0002-0786-7307]{G. Rodeghiero}
\affil{Max-Planck-Institut f\"ur Astronomie, K\"onigstuhl 17 D-69117 Heidelberg, Germany}

\author{M. Sawczuck}
\affiliation{Max-Planck-Institut f\"ur Astronomie, K\"onigstuhl 17 D-69117 Heidelberg, Germany}

\author{J.-U. Pott}
\affiliation{Max-Planck-Institut f\"ur Astronomie, K\"onigstuhl 17 D-69117 Heidelberg, Germany}

\author{M. Gl\"uck}
\affiliation{Universit\"at Stuttgart, Institut f\"ur Systemdynamik, Waldburgstr. 19, 70563, Stuttgart, Germany}
\affiliation{Max-Planck-Institut f\"ur Astronomie, K\"onigstuhl 17 D-69117 Heidelberg, Germany}

\author{E. Biancalani}
\affiliation{Max-Planck-Institut f\"ur Astronomie, K\"onigstuhl 17 D-69117 Heidelberg, Germany}

\author{M. H\"aberle}
\affiliation{Max-Planck-Institut f\"ur Astronomie, K\"onigstuhl 17 D-69117 Heidelberg, Germany}

\author{H. Riechert}
\affiliation{Max-Planck-Institut f\"ur Astronomie, K\"onigstuhl 17 D-69117 Heidelberg, Germany}

\author{C. Pernechele}
\affiliation{INAF OPAD, Vicolo Osservatorio 5, 35122 Padova, Italy}

\author{V. Naranjo}
\affiliation{Max-Planck-Institut f\"ur Astronomie, K\"onigstuhl 17 D-69117 Heidelberg, Germany}

\author{J. Moreno-Ventas}
\affiliation{Max-Planck-Institut f\"ur Astronomie, K\"onigstuhl 17 D-69117 Heidelberg, Germany}

\author{P. Bizenberger}
\affiliation{Max-Planck-Institut f\"ur Astronomie, K\"onigstuhl 17 D-69117 Heidelberg, Germany}

\author{L. Lessio}
\affiliation{INAF OPAD, Vicolo Osservatorio 5, 35122 Padova, Italy}



\begin{abstract}

The achievement of $\mu$arcsec relative astrometry with ground-based, near infrared, extremely large telescopes requires a significant endeavour of calibration strategies. In this paper we address the removal of instrument optical distortions coming from the ELT first light instrument MICADO and its adaptive optics system MAORY by means of an astrometric calibration mask. The results of the test campaign on a prototype mask (scale 1:2) has probed the manufacturing precision down to $\sim$ 50nm/1mm scale, leading to a relative precision $\delta\sigma \sim 5e-5$. The assessed manufacturing precision indicates that an astrometric relative precision of $\delta\sigma \sim 5e-5 = \frac{50\mu as}{1 arcsec}$ is in principle achievable, disclosing  $\mu$arcsec near infrared astrometry behind an extremely large telescope. The impact of $\sim$ 10-100 nm error residuals on the mask pinholes position is tolerable at a calibration level as confirmed by ray tracing simulations of realistic MICADO distortion patterns affected by mid spatial frequencies residuals.  We demonstrated that the MICADO astrometric precision of 50 $\mu$as is achievable also in presence of a mid spatial frequencies pattern and manufacturing errors of the WAM by fitting the distorted WAM pattern seen through the instrument with a 10$^{th}$ order Legendre polynomial.      

\end{abstract}


\section{Introduction} 
\label{intro}

 The Multi-AO Imaging CAmera for Deep Observations (MICADO) will be one of the first light instruments of the Extremely Large Telescope (ELT) providing near infrared (0.8 - 2.4 $\mu$m) high-resolution imaging with a special target on relative astrometry, spectroscopy and coronagraphy \citep{Davies18}. MICADO is assisted by a Single-Conjugate Adaptive Optics (SCAO) system, that will provide a corrected Field of View (FoV) of 19" and by the Multi-conjugate Adaptive Optics RelaY (MAORY) that will enable diffraction limited observations over the whole instrument FoV = 53", by means of two post-focal deformable mirrors conjugated at about $\sim$5 km and $\sim$12 km height \citep{Diolaiti14}. Reaching 50 $\mu$as relative astrometric precision requires a paramount calibration effort that needs to tackle the problem from multiple sides. As reported by \citet{Schoeck16} and \citet{Trippe10}, the astrometric error budget of a ground-based instrument embraces instrumental, atmospheric and astronomical systematic errors.  \citet{Rodeghiero18} dealt with the astrometric errors originated in the ELT telescope that need to be necessarily calibrated on sky. This work analyzes the distortions coming from the post-focal instruments, MAORY and MICADO, that can be calibrated by means of calibration masks deployed at their entrance focal planes. 
MICADO will be equipped with a Warm Astrometric Mask (WAM) deployed at the MAORY entrance focal plane, and a Cold Astrometric Mask (CAM) deployed at the MICADO entrance focal plane inside the cryostat. This paper focuses on the manufacturing and qualification efforts for the WAM prototype. The concept of calibrating the optical distortions with a reference target mask has been proposed \citep{Trippe10} and used already by different instruments like GeMS/GSAOI \citep{Riechert18} and NIRC2 \citep{Service16}. The calibration strategy relies on the accurate knowledge of the mask pinholes position that is compared with the image of the mask taken with the instrument and affected by the optical distortions. Fitting a series of different order polynomials on the image of the pinholes pattern leads to derive a 2D map over the FoV of the intrinsic distortion of the instrument. The polynomial fitting is normally done using different basis like Cartesian, Chebyshev and Legendre polynoms \citep{Riechert18}. The optical distortion coefficient scales with the third power of the field angular coordinate $\sim \theta^3$, so most of the distortions are removed already with a 3$^{rd}$ order polynom. The additional presence of mid spatial frequencies residuals at $\sim$ mm scale from the optics manufacturing process requires higher order polynomial like 5$^{th}$ or 9$^{th}$ that translates in a higher number of reference sources needed. In this context, a calibration mask ensures a dense grid of reference points that lead to sample completely the distortion pattern. To achieve the MICADO astrometric precision, $\sigma =$ 50 $\mu$as over 1 arcsec FoV, the calibration mask pattern needs to be known with a relative precision of $\delta\sigma \sim \frac{\sigma}{FoV} = 5e-5$ that for a pair of pinholes 1 mm apart translates in $\delta_{1mm} = 50$ nm. There are two options to tackle this extreme astrometric precision: an ultra stable and controlled manufacturing process of the calibration mask or an extremely precise metrology system for the qualification of the mask. In the case of the WAM prototype the approach chosen is hybrid; a significant effort was put in the identification of an adequate manufacturer especially in relation to the size of the WAM (200 mm x 200 mm), and a dedicated interferometric setup was developed to measure a sample of pinhole pairs separation on the mask. The paper summarizes the development process of the WAM prototype (scale 1:2) in the perspective of the MICADO Preliminary Design Review (PDR, Nov. 2018) starting from the concept of astrometric calibration with a mask (chapter \ref{concept}), the experimental setup built for the WAM characterization (chapter \ref{setup}), the results (chapter \ref{cap_result}) assessed in the test campaign and the impact of the residual manufacturing errors on the MICADO astrometric calibration (chapter \ref{cap_astrometry}).

\section{Warm Astrometric Mask calibration concept} \label{concept}

\begin{figure}
\epsscale{0.8}
\plotone{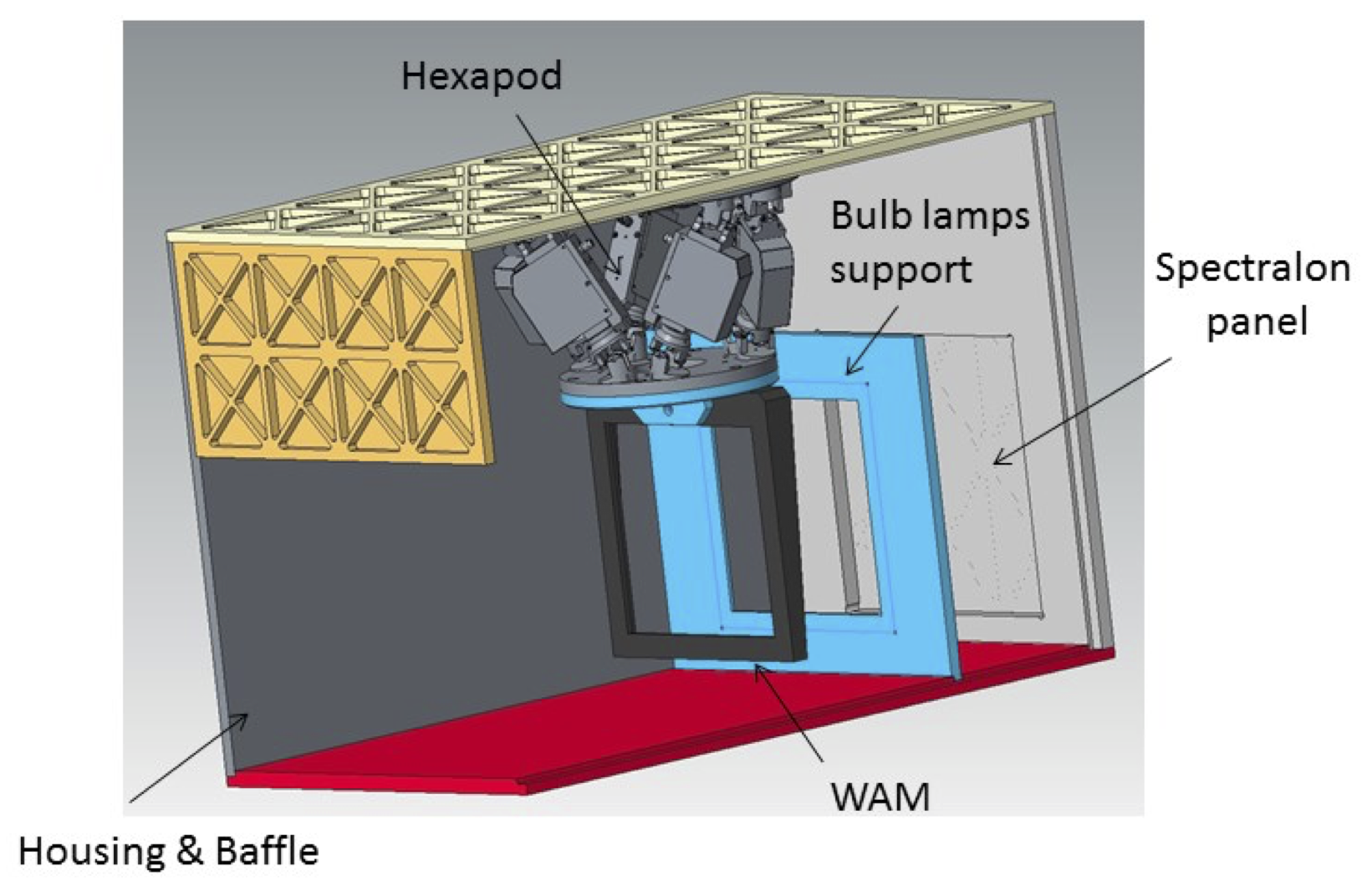}
\caption{Overview of the main astrometric calibration unit components with the WAM mounted on a hexapod for dithering, alignment and refocus of the mask. The mask is retro-illuminated with small miniaturized tungsten lamps and a Spectralon panel.\label{acu}}
\end{figure}

The MICADO astrometric calibration concept relies on two main pillars: the calibration of the telescope that is performed in sky observing reference star fields and self-calibration techniques, and the calibration of the post-focal instruments (MAORY and MICADO) that relies on the astrometric calibration masks \citep{Pott18}. The astrometric calibration masks are two: warm (WAM) and cold (CAM). The former, placed at the entrance focal plane of MAORY \textit{sees} both the MAORY and MICADO distortions, the latter, placed in the MICADO cryostat, sees only the MICADO optical distortions. The astrometric calibration unit contains the WAM that can be positioned with sub-micron accuracy along its three axes with an active hexapod to increase the distortion spatial sampling and map intra and inter-pixel sensitivity \citep{Rodeghiero18_2}. The WAM illumination is achieved with an array of 3x3 miniaturized tungsten lamps positioned behind the WAM and emitting light towards a Spectralon panel that reflects and diffuses the light in the direction of the mask as shown in Fig. \ref{acu}. The diffused light enters for transmission the WAM and makes diffraction in correspondence of its pinholes. The pinholes act as an array of artificial point-like sources that are reimaged by the post- focal instruments (MAORY and MICADO). Knowing with high accuracy the position of the WAM point-like sources leads to estimate and remove the geometric distortion of the instruments by comparing the nominal (mask) and distorted (image) point-like sources patterns.

The trade-off for the selection of the WAM substrate and production technology concentrates on two different material families and manufacturing technologies: (i) ceramic and glass substrate treated with coating photo-lithography and (ii) metal substrate manufactured with laser drilling. Based on the experience of the MUSE experiment \citep{Kelz12}, there are several cons and pros for both the technologies as reported in Table \ref{table1}.

\begin{table*}
\begin{center}
\begin{tabular}{p{5.5cm}*{2}{c}} \hline
\textbf{Lithography ceramic/glass} & \textbf{Laser drilling  metallic plate} \\
\hline 
Etching of Chrome coating & Laser drilling, micro-drilling (electro-erosion) \\
Residual background 10$^{-3}$-10$^{-4}$& No background \\
Polished window & Flatness TBE\\
Low thermal expansion & High thermal expansion \\
Low thermal conductivity & High thermal conductivity\\
Higher accuracy pitch and diameter & Lower accuracy pitch and diameter\\
Lower accuracy PSF photometry & Higher accuracy PSF photometry \\
\hline
\end{tabular}
\caption{Comparison of the main cons and pros of the two envisaged technologies and substrates evaluated for the WAM manufacturing. \label{table1}}
 \end{center}
 \end{table*}

\begin{figure}
\plotone{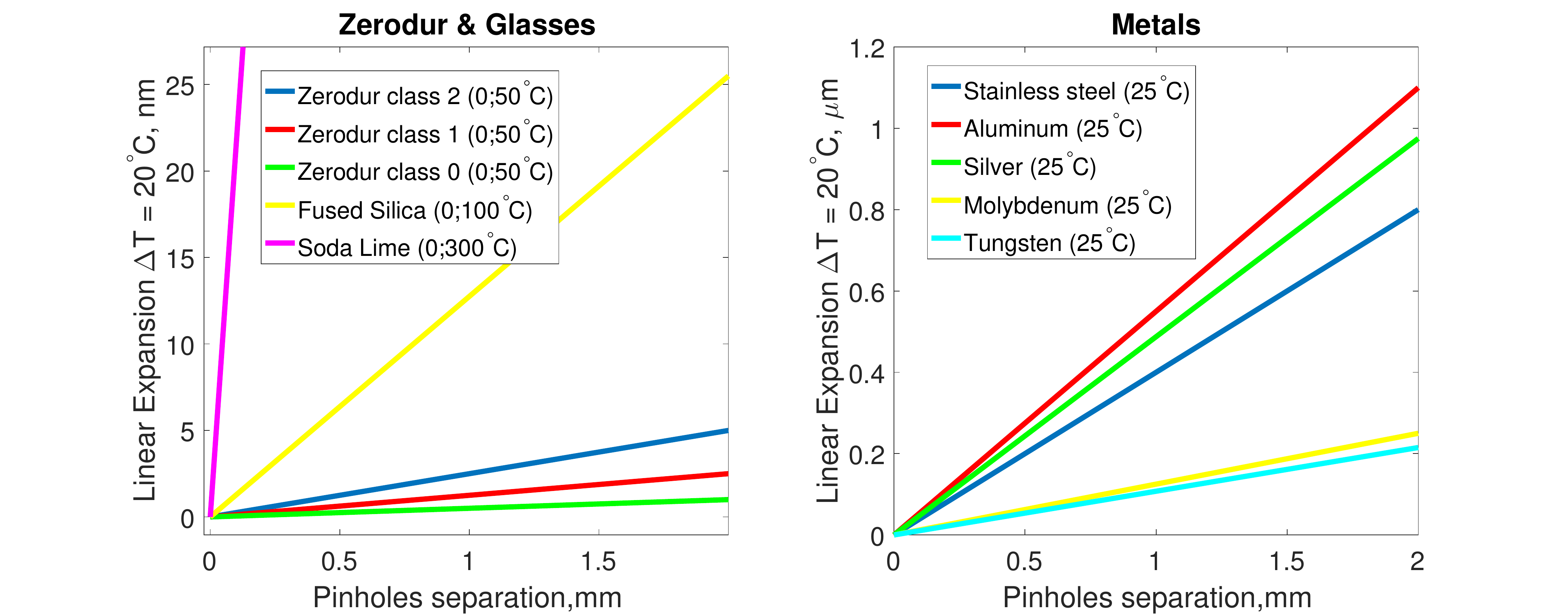}
\caption{Left: linear expansion of a 2 mm substrate for different Zerodur substrates and glass materials within a $\Delta T$ = 20$^\circ$C. The metals substrates (right) have systematically higher CTE and expansion lengths that prevent their use for the WAM. \label{zerodur}}
\end{figure}

The main driver for the substrate selection is the thermo-mechanical stability of the WAM in the required operational range (0 $^\circ$C to + 20$^\circ$C). Any linear/volumetric expansion or contraction induced on the WAM challenges the reliability of the astrometric calibration thus requiring the use of an ultra-stable material for the substrate. The linear thermal expansion, $\Delta l = \alpha\Delta Tl$, with $\Delta T$ temperature variation, $l$ linear size of bulk considered and $\alpha$ Coefficient of Thermal Emission (CTE) of the material, estimates the linear size variations (expansion or contraction) for the substrate. As shown in Figure \ref{zerodur}, the expected substrate length variations, over a typical pinholes separation of 2 mm, rules out all the metallic substrates and soda lime glass. The most reliable material in the perspective of a day-night calibration within a temperature variation range of $\sim$ 20$^\circ$C is therefore Zerodur with average relative expansion factor of $\Delta l/l\sim0.05 \cdot 10^{-6}$ . Interesting studies about the Zerodur thermal behaviour and hysteresis \citep{Jedamzik10} indicate that a Zerodur substrate class 0-1 is the most stable material for the typical night-day temperature cycle profile at Armazones (ELT construction site) without the need of any thermalization of the WAM.

\begin{figure}[!h]
\plottwo{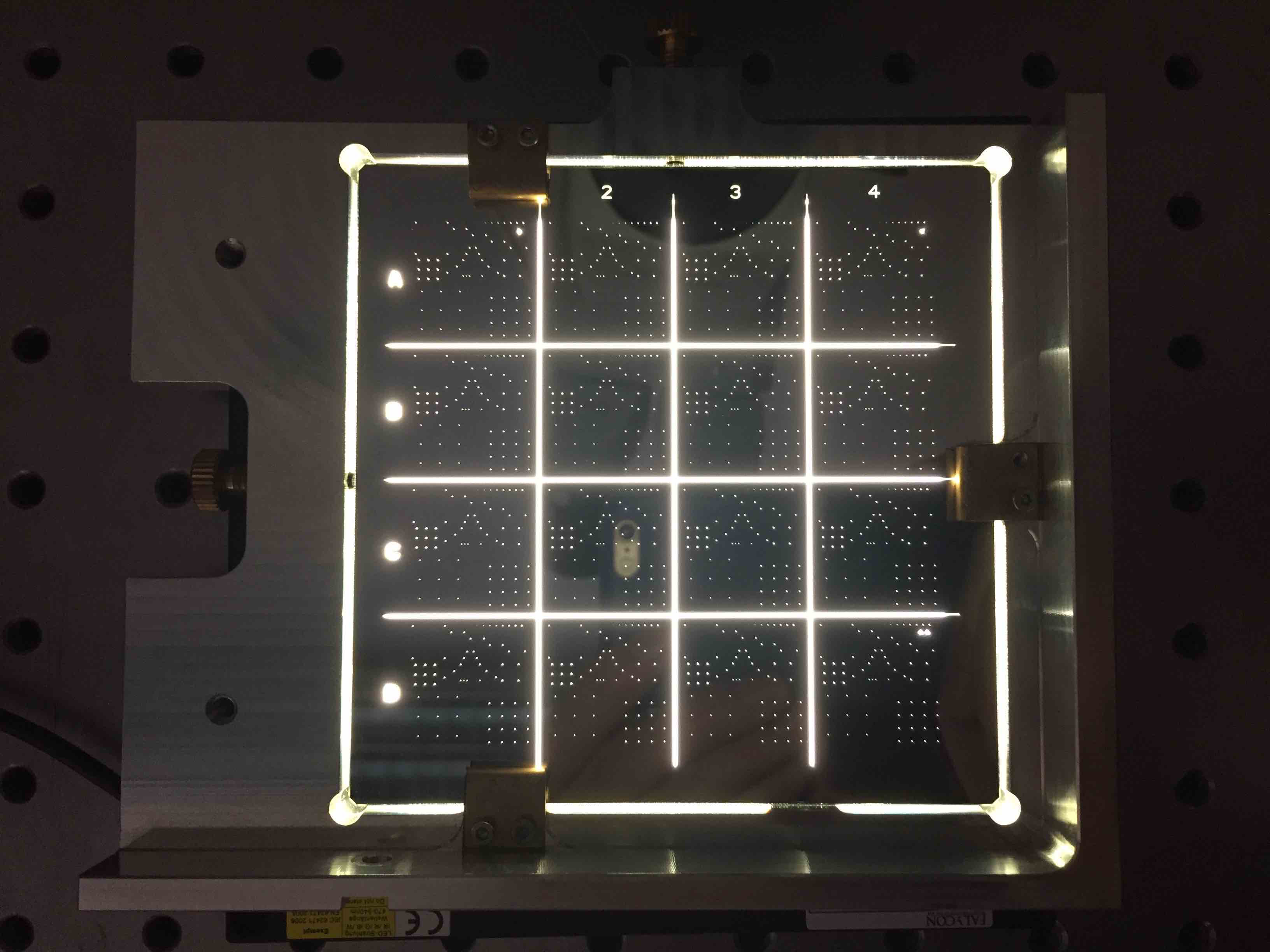}{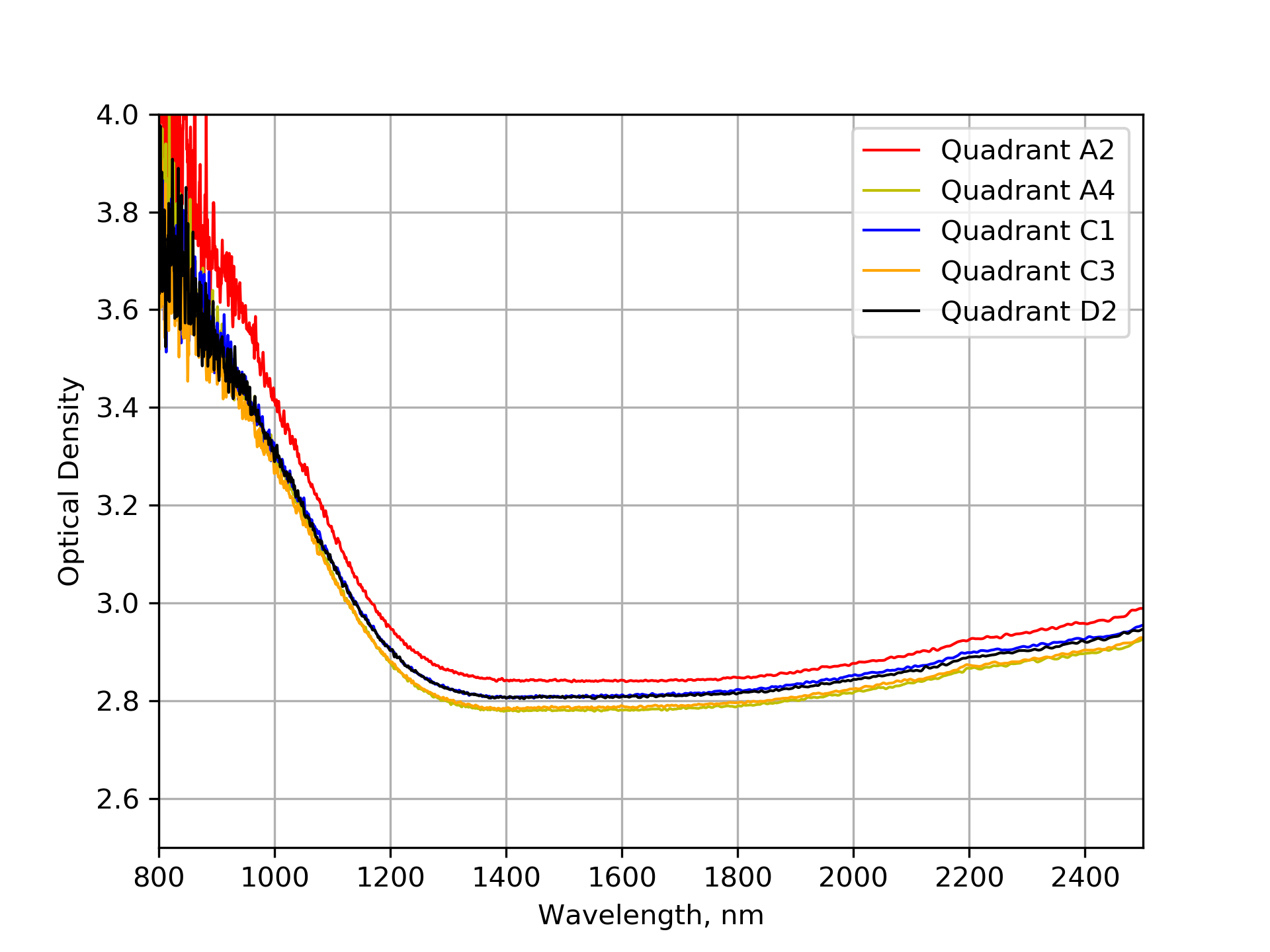}
\caption{Left: picture of the WAM prototype tested at MPIA backside illuminated showing the 16 twin quadrants (A..D x 1..4). Right: optical density of the WAM Chrome coating in five different quadrants measured with a Nicolet iS50 FTIR. The range of interest for MICADO astrometry is H and K band, between 1500 nm and 2400 nm characterized by an optical density of $\sim$2.8-2.9. \label{coating_wam}}
\end{figure}

The Zerodur substrate is coated with a brown Chrome layer of $\sim$ 150 nm thickness that results in an optical density of 2.8-3 at NIR wavelengths (Fig. \ref{coating_wam}). The production of the pinholes pattern is carried out in a series of operations: (i) the Zerodur substrate is coated with a sputtered brown Chrome layer (ii) a layer of resistant lacquer is deposited on top of the Chrome layer (ii) the coating is etched with a He-Cd UV laser in correspondence of the pinholes position by a photo-lithography process. The WAM prototype has been produced with a series of multiple pinholes pattern with different diameters (from 5 $\mu$m to 50 $\mu$m) and separation (from 1 mm to 8 mm) to facilitate the test campaign.

\section{Measuring at sub-micron scales} \label{setup}

\begin{figure}
\epsscale{0.5}
\plotone{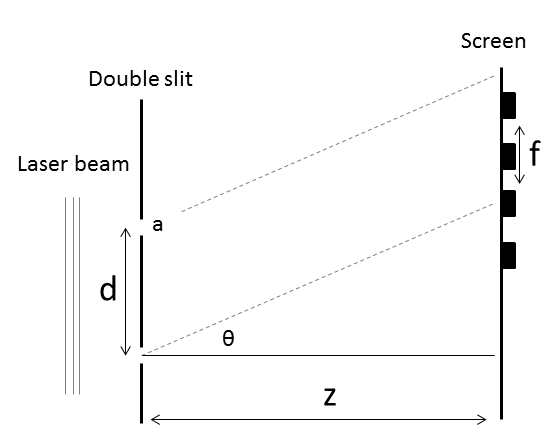}
\caption{Schematics of the Young's double slit experiment in relation to the parameters of Eq.\ref{young} The diameter of the pinhole is $a$  and it controls the amplitude of the diffraction pattern envelope from the single pinhole and it is of marginal interest for the current study. The observable of the experiment is the fringe separation $f$ that is used to estimate the pinhole separation $d$.\label{young_book}}
\end{figure}

The measurement of the pinholes pair separation on the WAM represents a major effort in the verification of the MICADO calibration scheme. As discussed in chapter \ref{concept}, the current strategy relies both on in-sky calibrations of the telescope distortions and on a calibration masks set for the detection of distortions in the instrument. In this perspective, the WAM, as built, pinholes coordinates become the reference field points for the astrometric calibration with the polynomial fit of the instrument distortion map. A major issue in the metrology of the WAM is posed by the desired relative precision, $\delta\sigma \sim 5e-5$, that rules out standard imaging objectives or microscopes that suffer from unavoidable intrinsic optical distortions that are very difficult to disentangle from possible manufacturing errors of the mask. A first attempt with a microscope, Olympus SZX12, has led to a systematic error of $\sim$ 60 $\mu$m over a pinhole pair separation of 1.5 mm (distortion $\sim$ 4 \%). 
To overcome this issue imposed by the direct imaging, we decided to setup a test bench based on the Young's double slit experiment. The test measures the separation $d$ between two pinholes (double slit) by observing the interference fringe separation $f$ created on a screen at distance $z$ when the mask is illuminated with a coherent light source of wavelength $\lambda$ as shown in Fig. \ref{young_book} and described by Eq. \ref{young} in the far-field approximation $z>>d$.  

\begin{equation}\label{young}
    d = \frac{z\lambda}{f}
\end{equation}

The great advantage of this setup is the absence of any optics (so distortion) between the mask and the detector. 

\begin{figure}
\epsscale{0.85}
\plotone{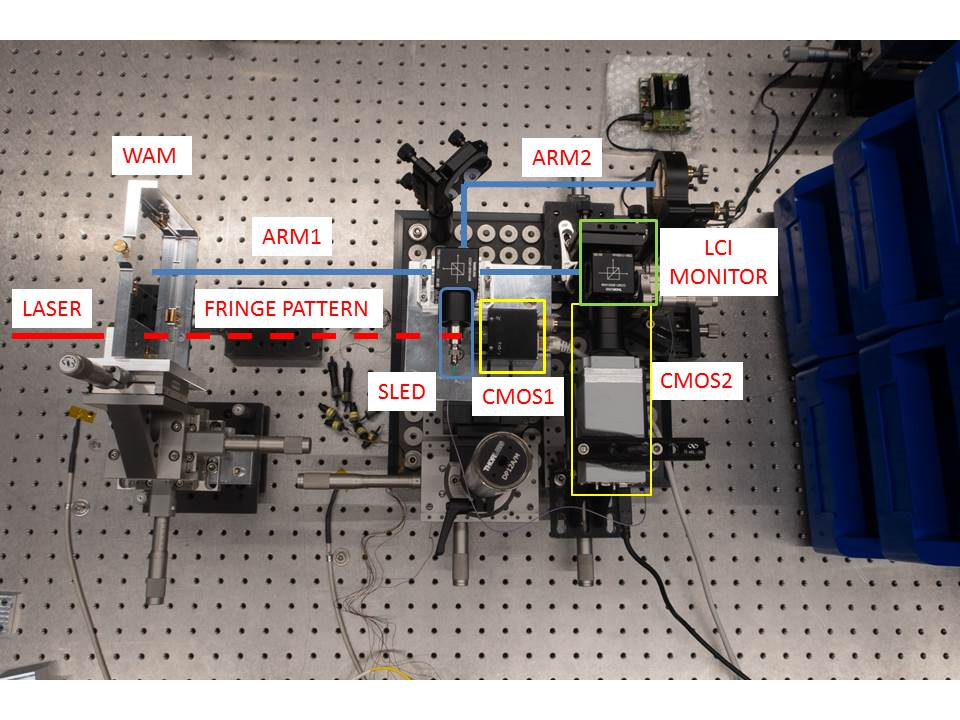}
\caption{Lab setup for the measurement of the WAM prototype pinholes separation. The setup is based on a Young's double slit experiment with a laser illuminating a pair of pinholes on the WAM and a detector collecting the fringes to estimate the fringes separation. A LCI Michelson interferometer monitors the separation between the WAM and the detector during the measurement. ARM1 is $z$ in eq. \ref{young} as the WAM is used like mirror of the LCI interferometer, while ARM2 is assumed to be invariant being mounted on a carbon fiber bench with ultra-low thermal expansion. CMOS1 records the Young's fringes, while CMOS2 tracks the SLED fringe shift in the LCI.\label{LCI_setup}}
\end{figure}

The interference pattern at the screen is described by the Fraunhofer diffraction pattern equation (\ref{fraunhofer}):

\begin{equation}\label{fraunhofer}
    I(\theta) = 4I_0\left(\frac{\sin \beta}{\beta}\right)^2\cos^2{\alpha}
\end{equation}

the low-spatial frequency envelope coming from the diffraction through the single pinhole is described by the $sinc$ function while the fast oscillating component associated with the fringes is given by the squared cosine. The coefficient of the cosine, $\alpha = \frac{2\pi(x - c)}{f}$, contains the fringe separation $f$ that is the main observable of the experiment.  The practical implementation of the Young's experiment in shown in Fig. \ref{LCI_setup}: a He-Ne laser is projected onto a pair of pinholes on the WAM and the interference pattern is recorded with CMOS1 (Prosilica GC1600H), sensitive area 7.13 mm x 5.37 mm - 4.4 $\mu$m pixel pitch, at a distance $z\sim$ 300 mm (ARM 1 in Fig. \ref{LCI_setup}).  

\begin{figure}
\plottwo{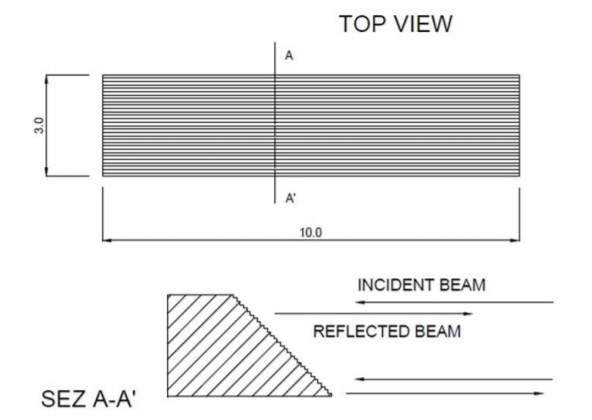}{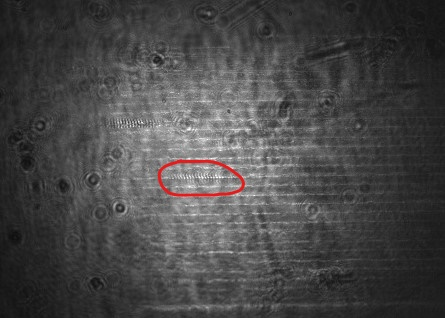}
\caption{Left: stepped mirror geometry \citep{Pernechele17}, 30 steps each corresponding to an optical path difference of 100 $\mu$m. Right: image of the stepped mirror taken with CMOS2 showing the LCI fringes (red circle). A second static fringe pattern, noteless, is visible on the top left of the stepped mirror that is the interference between a relay mirror (to reimage the stepped with CMOS2) and the stepped mirror. \label{stepped}}
\end{figure}

\begin{figure}[!h]
\plottwo{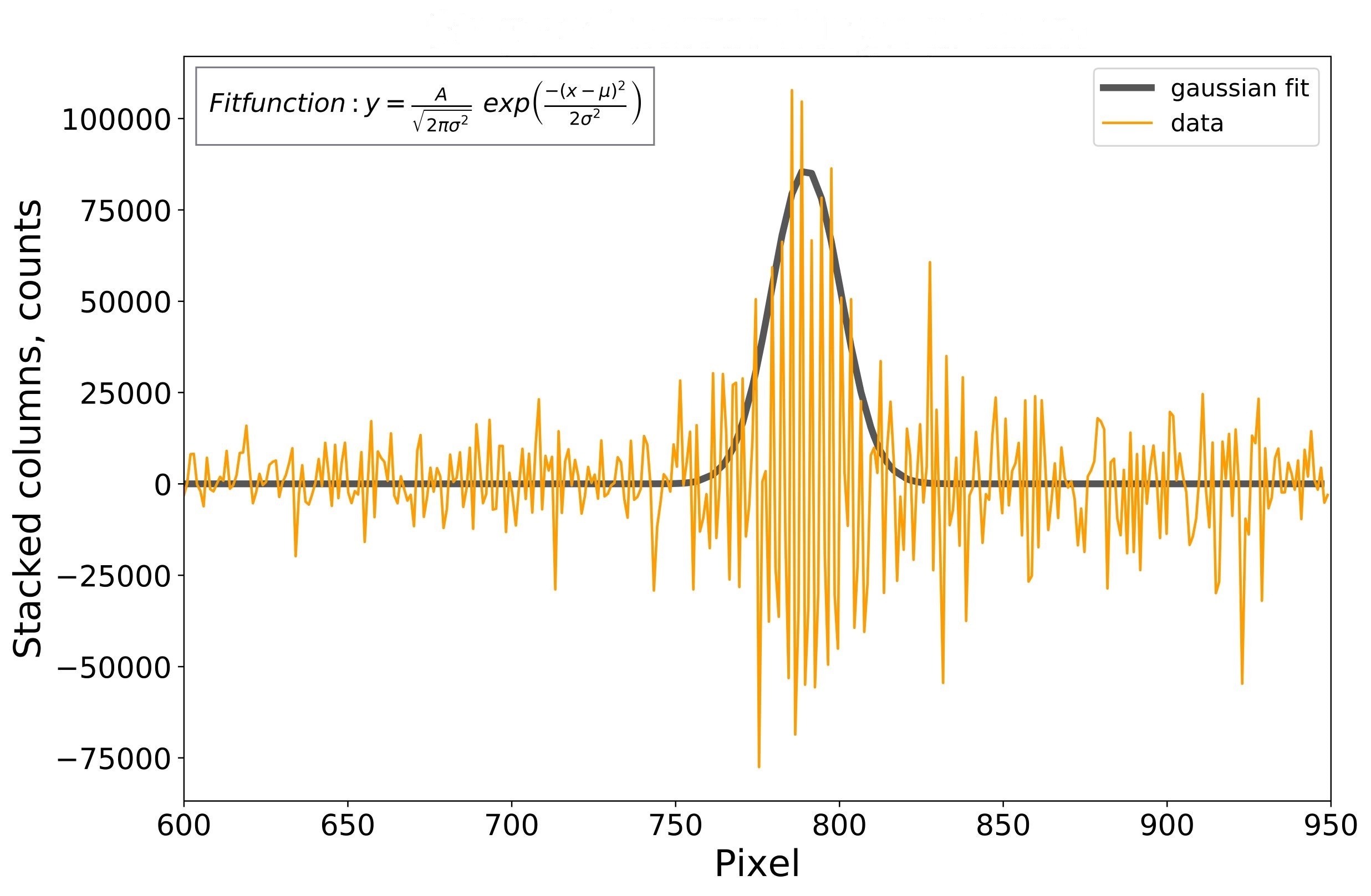}{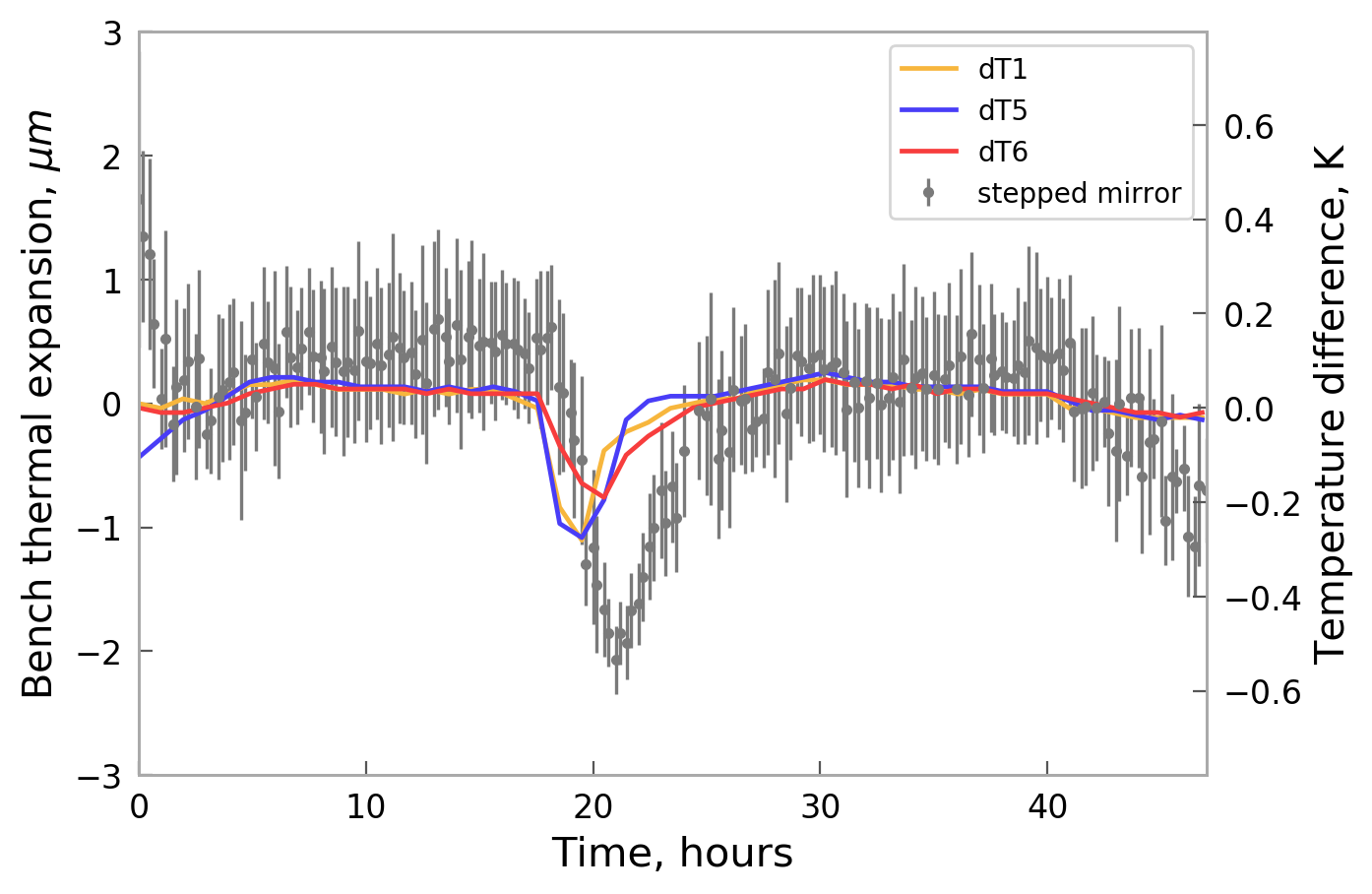}
\caption{Left: Gaussian fit of the LCI fringe pattern whose centroid provides a measurement of the WAM-to-detector baseline variation $\delta z$. Right: the LCI setup allows to monitor the optical bench diurnal cycle of thermal expansion/contraction showing a good agreement with a network of temperature sensors on the bench (color lines).   \label{centroid_sled}}
\end{figure}

A significant effort needs to be put in the metrology and monitoring of the relative separation $z$ between the WAM and the detector: to measure any relative displacement between the two objects at sub-micron level, a Michelson Low Coherence Interferometer (LCI) is installed in parallel to the Young's setup. The LCI technique has several applications and it is commonly used for measuring 3D surface morphology at sub-micrometric scale, thermo-elastic expansion of structures \citep{Wyant02}  and optical substrate thickness measurements \citep{Pernechele17}. This non-contact technique does not provide an absolute distance measurement of $z$ instead it measures at a submicron scale the variation $\Delta z$ arising from thermo-mechanical displacement of the WAM with respect to the detector. The light source for the LCI interferometer is a super-luminescent diode (SLED), model EXS210036-01, that has a small optical coherence length (12 $\mu$m) and a narrow bandwidth (25 nm) centered at $\lambda =$ 820 nm. The first arm of the LCI interferometer (ARM1) corresponds to $z$ (Fig. \ref{LCI_setup}), the second arm (ARM2) is placed on a carbon fiber bench with ultra low thermal expansion assumed to be length-invariant. The light from the two arms is combined in a beam splitter and projected on a stepped mirror (LCI monitor, Fig. \ref{LCI_setup}) that is reimaged by the camera CMOS2 (Pixelink PL-B957U). Being the SLED coherence length very short, the interference pattern from the LCI is localized in a small portion of a stepped mirror (example in Fig. \ref{stepped}) leading to easily fit a Gaussian profile on the fringe envelope and estimating its centroid as shown in Fig. \ref{centroid_sled}-left. As shown in Fig. \ref{stepped}, the stepped mirror is a polished optical ladder of 30 steps (100 $\mu$m per step) that provides a measurement dynamic range of 3 mm along the $z$ coordinate; e.g. a $\delta z$ = 100 $\mu$m on ARM1 translates in a jump of the SLED fringe from one step to another. To measure the absolute length of $z$ at a micron level we used a Zeiss ScanMax-IPX 3D Coordinate Measuring Machine (3D-CMM) by which we can also estimate the initial WAM and detector relative orientation touching the two objects with a small probe in different points. The 3D-CMM measurement sets also the zero-point on the stepped mirror, a given $z_0$ value is associated to a certain position of the LCI pattern on the stepped mirror. Any subsequent displacement or wobble of the WAM with a component  along the z coordinate translates in a motion of the LCI fringes $z_0+\delta z$ that does not require the 3D-CMM to be monitored. An interesting example of the potentialities of the LCI technique is shown in Fig. \ref{centroid_sled}-right, where the circadian thermal expansion and contraction of the optical bench (ARM1) is tracked by the LCI fringes motion and compared to the temperature measurements of bench. 
The light source to study the Young's fringes comes from an He-Ne laser ($\lambda$ = 632.816 nm), spatially filtered by two orthogonal slits, used to illuminate and isolate a specific pair of pinholes on the WAM. The WAM is aligned to the LCI interferometer by means of a tip-tilt and rotation stage and different pinhole pairs are engaged in the measurement using a three axes linear stage that holds the mask. The three axes linear stage, Newport 462-XYZ-M, with $<$100 $\mu$rad angular deviation (wobble) has an axial run-out $<$ 2 $\mu$m Peak to Valley (PV) over 25 mm range.
Two axes out of three are motorized with a stepper motor with 0.1 $\mu$m minimum incremental step. The pre-measurement sequence for a typical Young's fringes measurement foresees: (i) aligning the WAM to the LCI interferometer and equalize ARM1 and ARM2 within the stepped mirror dynamic range, (ii) selecting a certain quadrant and pinholes pair on the WAM, (iii) calibrating the stepped mirror using the motorized stage to move the WAM along $z$, (iv) measuring $z$ with the 3D-CMM.

\section{Test campaign results}
\label{cap_result}

\begin{figure}
\plottwo{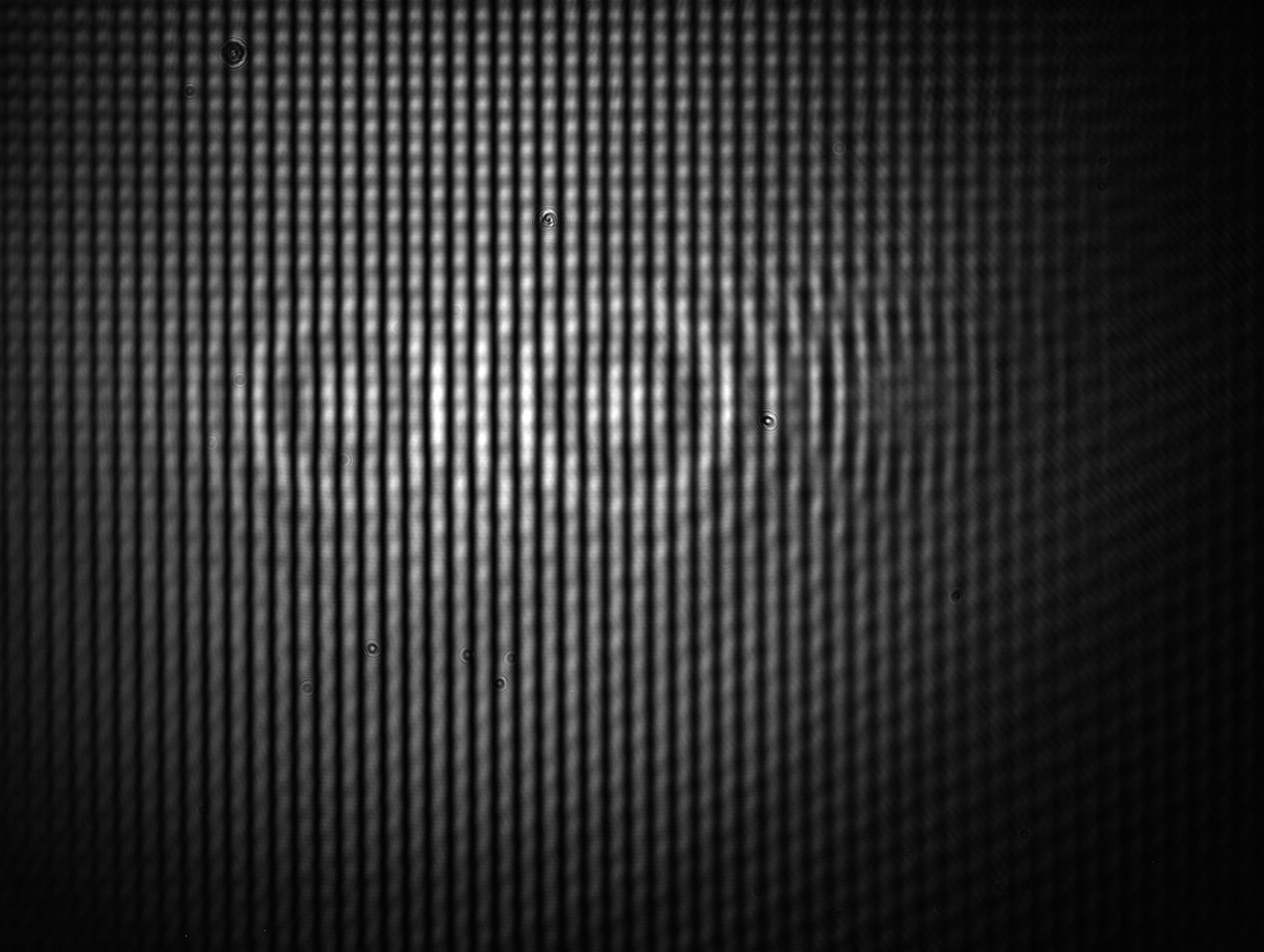}{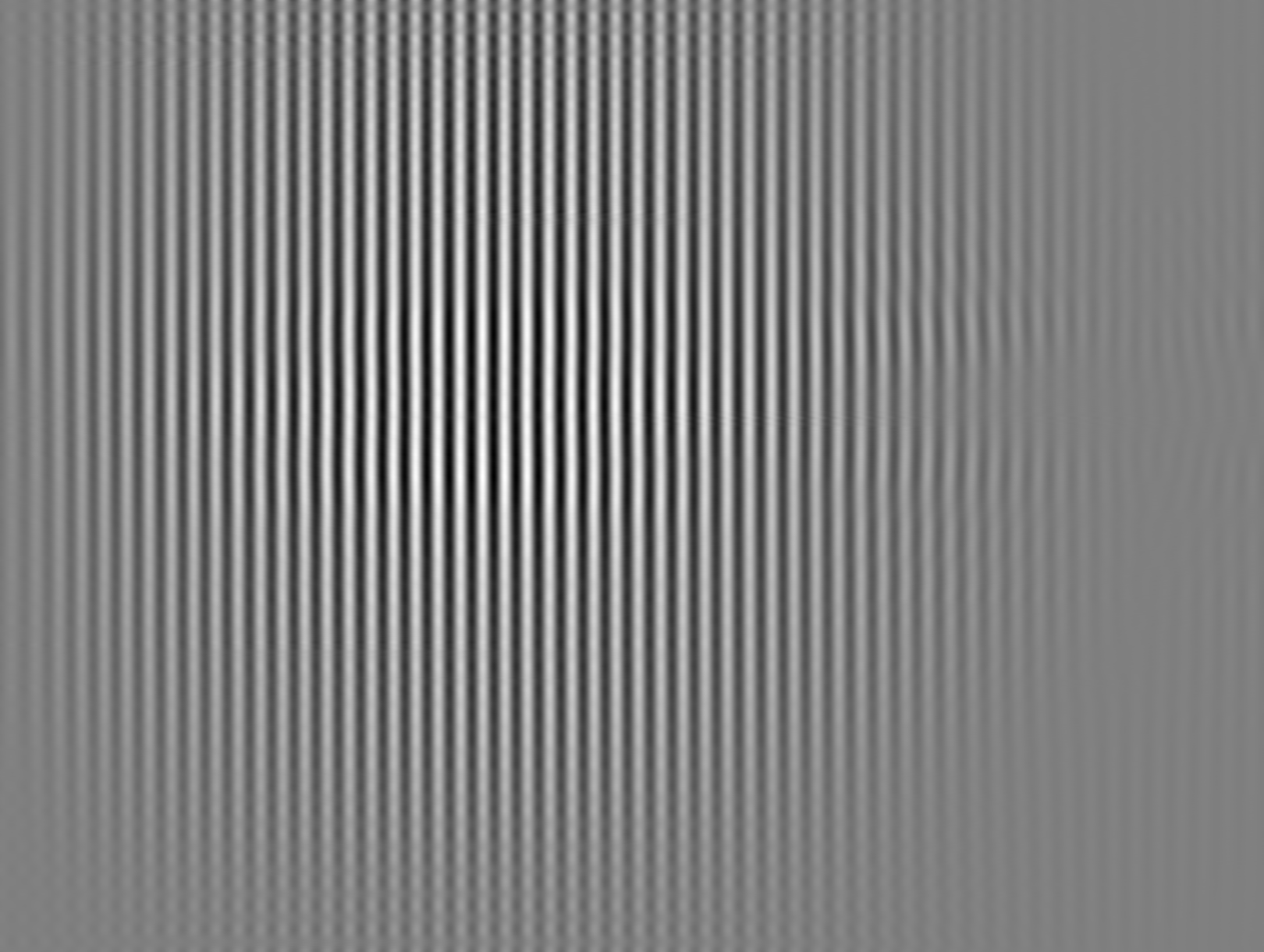}
\caption{Young's fringes obtained from a 50 $\mu$m pinholes pair separated by 1.5 mm (left raw) and after band-pass filtering (right). \label{raw_filt}}
\end{figure}

\begin{figure}
\epsscale{0.5}
\plotone{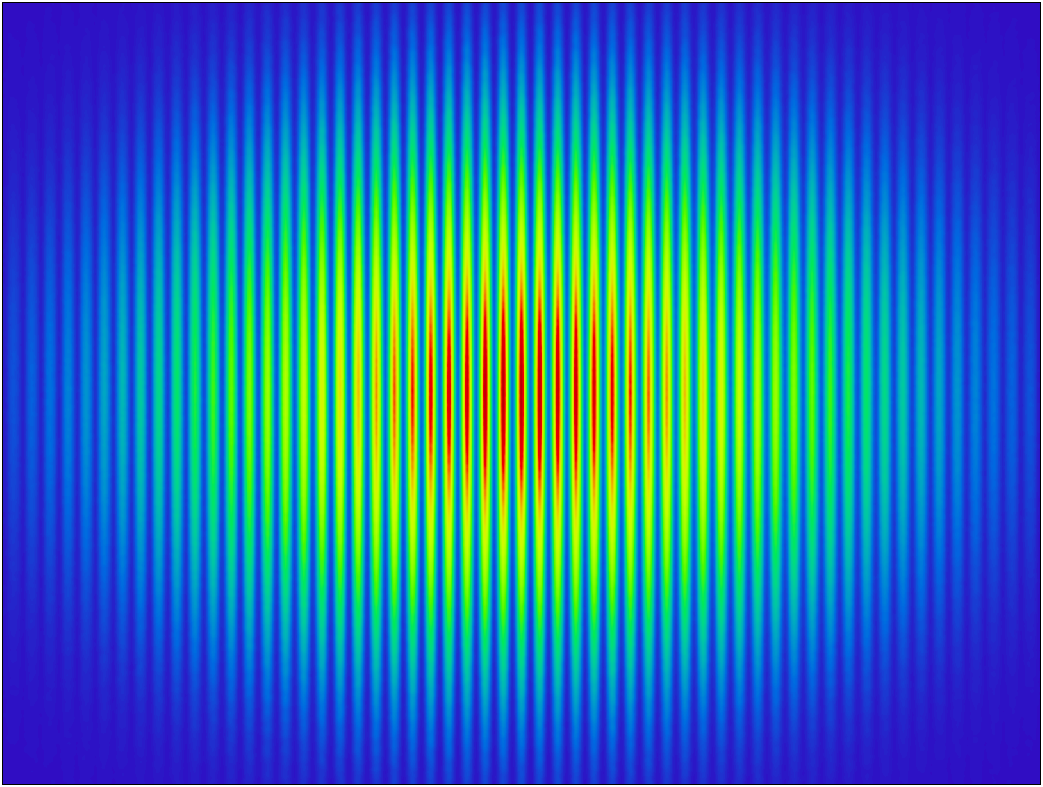}
\caption{Zemax simulation of the Young's fringes from two 50 $\mu$m pinholes 1.5 mm apart using a virtual detector with the CMOS1 specifications. The rings artifacts affecting the raw image (\ref{raw_filt}) are not visible while the first minimum (Eq. \ref{eq_single_pin}) of the single pinhole diffraction pattern at edge of the detector is clearly visible.\label{fringe_zmx}}
\end{figure}

\begin{figure}
\epsscale{0.7}
\plotone{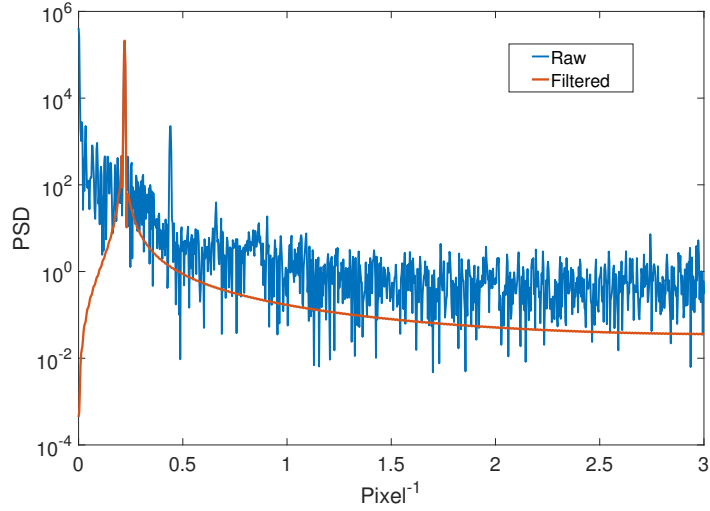}
\caption{PSD analysis of the fringe pattern coming from two 50 $\mu$m pinholes before (blue) and after (red) bandpass filtering. The filtered signal isolates the harmonic associated with the Young's fringes.\label{fringe_PSD}}
\end{figure}

The test campaign on the WAM prototype assessed a total of 14 pinhole pairs in three different quadrants of the mask, spanning a range of $\sim$ 80 mm on the mask surface. The analysis was restricted to pinholes pairs 1.5 mm apart with a 50 $\mu$m diameter to maximize the signal to noise ratio (SNR). Examples of Young's fringes measured using the setup in Fig. \ref{LCI_setup} are shown in Fig. \ref{raw_filt}. The Fraunhofer diffraction pattern (\ref{fraunhofer}) is composed by a low spatial frequency envelope associated with the diameter of the single pinhole $a$ and a high frequency fringe pattern due to the interference between the two apertures. The size of the light pool $y$ at the first minimum (m = 1) associated to the single pinhole is described by Eq. \ref{eq_single_pin}. For the current setup, $z$ $\sim$ 297 mm and $d$ = 1.5 mm, the diameter of the pinhole diffraction envelope is $2y$ = 7.52 mm; the fading in correspondence of the first minimum at the edge of the detector is visible both in the real image \ref{raw_filt}-left and the simulated one (Fig. \ref{fringe_zmx}). The raw frame Fig. \ref{raw_filt} is affected by some smaller rings of unknown origin likely attributable to the laser beam, but of no interest for the fringes spacing measurement. To improve the data quality and cosmetic, the raw images were band-pass filtered to isolate in the power spectrum the harmonic associated to the fringes spacing $f$ (Fig. \ref{fringe_PSD}). Being the fringe pattern oriented vertically with respect to the image frame, a one-dimensional digital band-pass filter is designed and applied to each row using by an infinite impulse response (IIR) band-pass filter. Hence, the minimum filter order can be achieved and the transient filter effects are small. Furthermore, the transient effects are reduced by zero padding. Since a digital filter is used, the image can be filtered without a phase shift (zero-phase filtering). The filter parameters are determined by using the theory for a double slit. Thus, we obtain the spatial frequency of the fringes, which is the center frequency $f_0$ of the band pass filter. The cut-off frequencies are given by $f_0 \pm \Delta f$. An example of a filtered image is shown in Fig. \ref{raw_filt}-right.

\begin{equation}\label{eq_single_pin}
    y = \frac{mz\lambda}{d}
\end{equation}

\begin{figure}
\plottwo{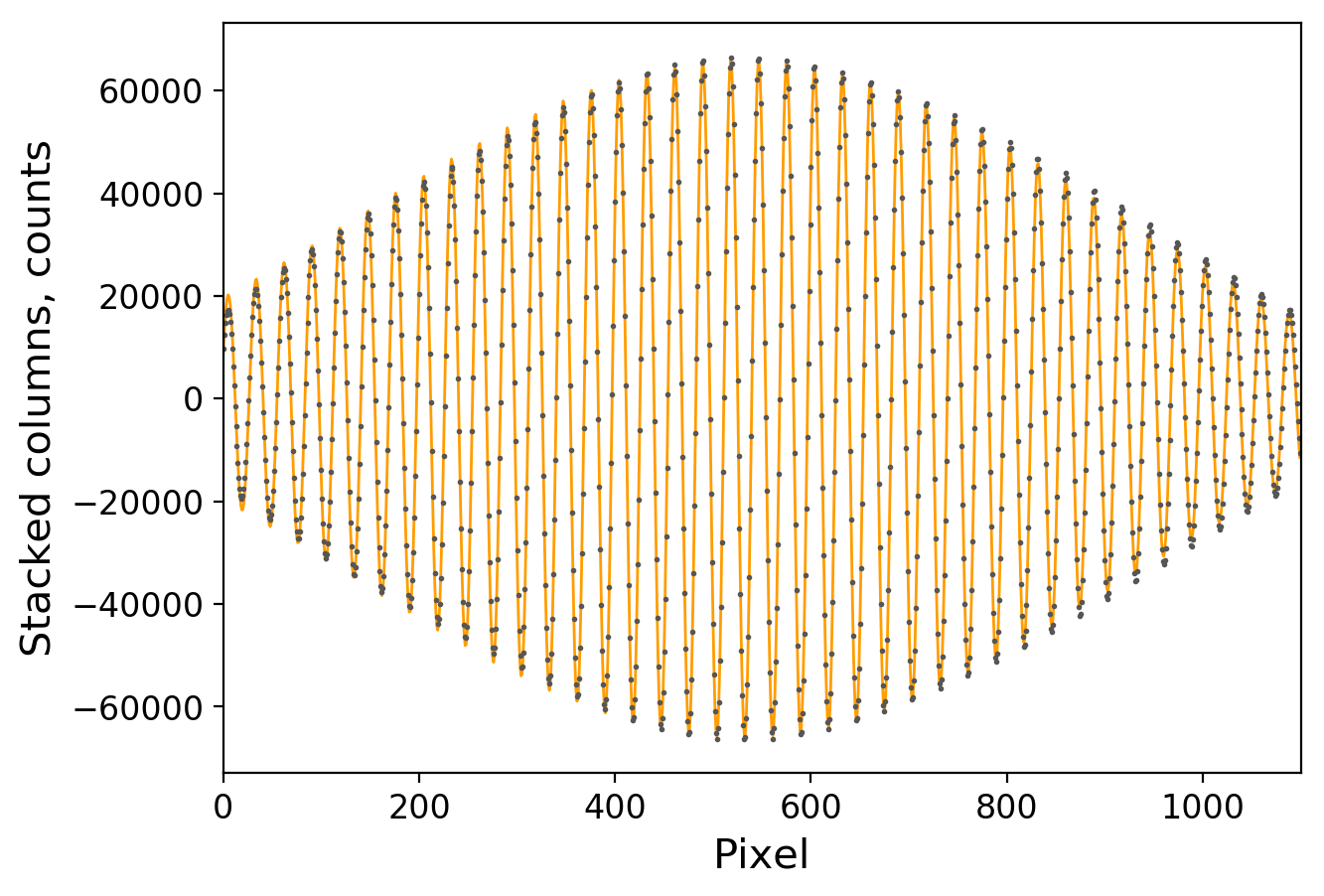}{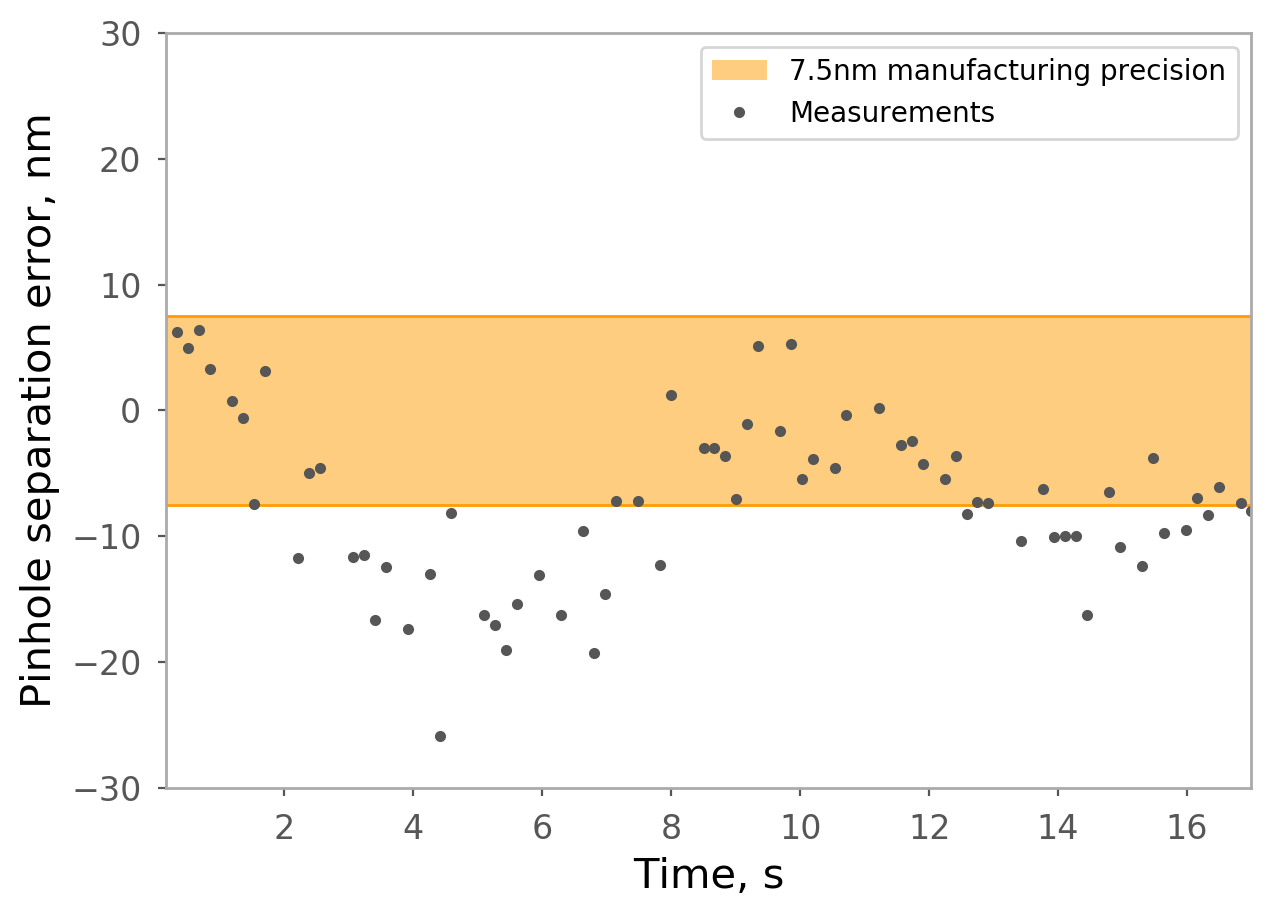}
\caption{Left: stacked 1D slice (black points) of the Young's fringes obtained from a 1.5 mm-50 $\mu$m pinholes and fitted Fraunhofer diffraction pattern (orange line). The extraction of only the fringe spatial frequency leads to remove the mean value of the signal and therefore the obtained intensity values are partially  negative. Right: Residuals errors on $d$ measurement from a pinholes pair for a series of 100 frames. The orange band shows the expected manufacturing precision from Eq. \ref{imt_eq}. \label{cos_100_frames}}
\end{figure}

After the frames filtering, a region in the core of the image is extracted, where the fringes are sharp and the SNR is high. The columns of the image subset are stacked to obtain a 1D pixel slice collecting all the counts of the region of interest and subsequently fitted with the Fraunhofer diffraction pattern (Eq. \ref{fraunhofer}) as shown in Fig. \ref{cos_100_frames}-left. The fit retrieves an estimate of $\alpha (f)$ that leads to derive the fringes separation and the pinholes separation $d$. To increase the data sample statistics a series of 100 frames is taken for each pinholes pair. Slow drift and amplitude variations are observed in the temporal evolution of $\Delta d = d_{meas} - d_{exp}$ (measured - expected) as shown in Fig. \ref{cos_100_frames}-right. The origin of these systematic errors is likely associated to the vibration state of the WAM for the small-amplitude fast (sub-sec) component and to a mechanical relaxation of the WAM support for the larger and slower ($\sim$ sec) amplitude drifts. The selection of the pinholes pair involves a manual stage and a motorized one. A post-processing measurement with a vibrometer, Polytec OFV-505, of the fast small-amplitude, unavoidable, drift finds the WAM in a omni-directional vibration state with an amplitude of $\sim$ 3 $\mu$m that translates into a $\Delta d\sim$ 15 nm (PV) in agreement with the small scale dispersion of the points in Fig. \ref{cos_100_frames}; the slower and larger, $\sim$ 30-80 nm (PV), perturbation is more variable from pair to pair being likely driven by the hardly repeatable action of the operator that moves the WAM with the manual stage. We observed that waiting for a few minutes after the pinhole pair selection before starting the acquisition of the frames leads to smaller systematic error on $\Delta d$ (PV $\sim$ 30 nm, Fig. \ref{cos_100_frames}-right).

\begin{figure}
\epsscale{0.75}
\plotone{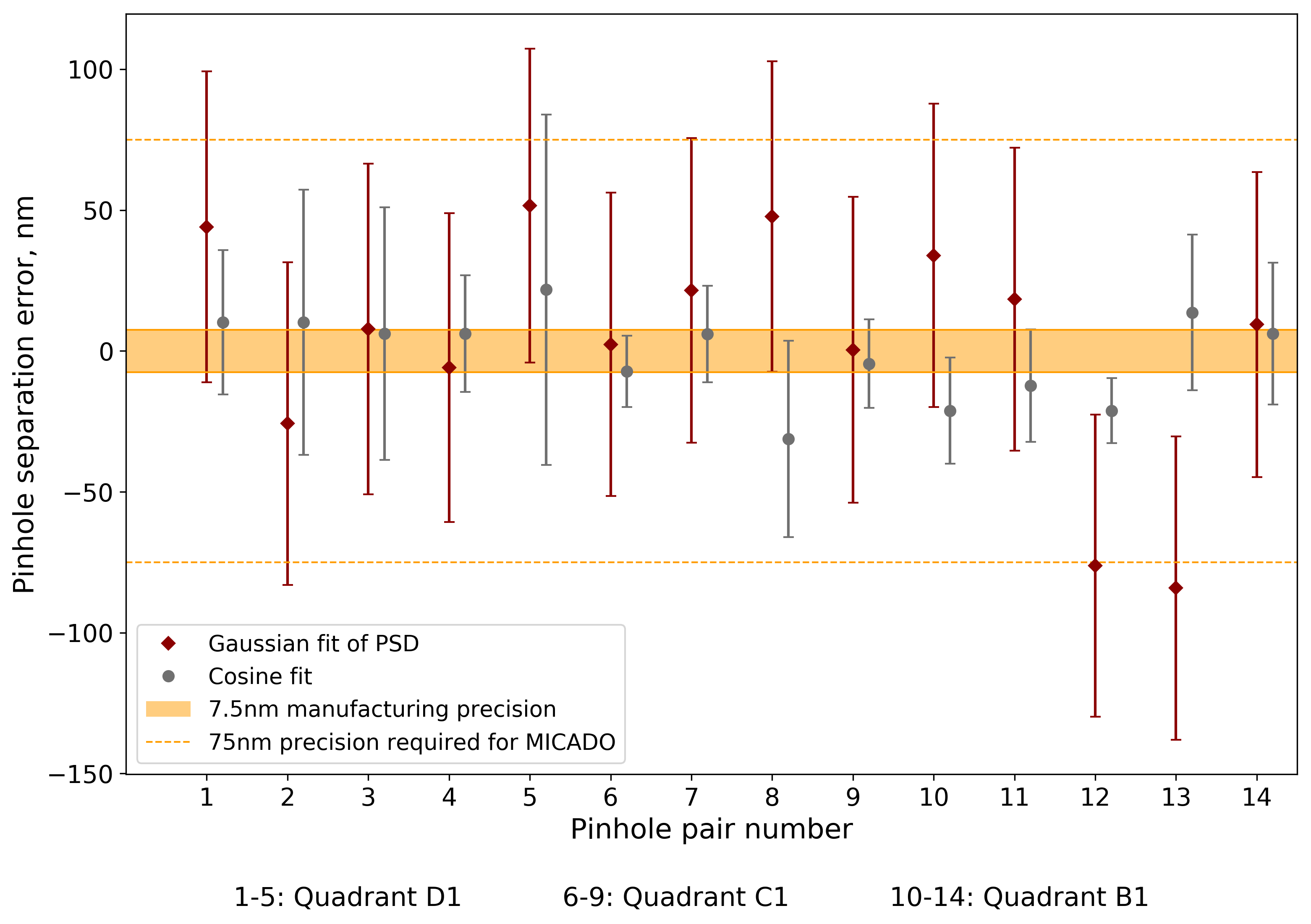}
\caption{Median of value of the pinholes separation residual error $\Delta d = d_{meas} - d_{exp}$ measured in 100 frames for 14 different pairs from three WAM quadrants. The orange band represents the expected manufacturing precision of the mask (Eq. \ref{imt_eq}) while dashed line shows the maximum tolerable pinholes separation error corresponding to an astrometric relative precision of 50 $\mu$as over 1 arcsec FoV. The gray dots are derived fitting the Fraunhofer diffraction pattern, the red dots from the PSD peak fitting.\label{results_plot}}
\end{figure}

An overview plot of the test campaign results from 14 pinhole pairs distributed over a range of 80 mm on the WAM is reported in Fig. \ref{results_plot}. The plot reports the estimate of $\Delta d$ using two different methods: the fit of the fringe pattern with the Fraunhofer diffraction law (gray points) and the Gaussian fit and centroiding of the PSD peak associated to the fringes harmonic in the Fourier space (red points).

The two techniques give comparable results and have comparable data points dispersion. The amplitude of the error-bars is driven by the error on the baseline $z$ (Eq. \ref{erro_budget}) and on the fringe separation $f$ (fringe distortions and fit uncertainty).

\begin{equation}\label{erro_budget}
    \sigma_z = \sqrt{\Delta z^2_{det} + \Delta z^2_{CMM} + \Delta z^2_{ori} + \Delta z^2_{LCI}} 
\end{equation}

The major contributors to $\sigma_z$ are the uncertainty on the detector focal plane position with respect to its enclosure $\Delta z_{det}=\pm 10 \mu$m, the error of the 3D CMM and the WAM orientation $\Delta z_{CMM}=\Delta z_{ori}\pm 5 \mu$m. For the current setup the average uncertainty is $\sigma_z \sim$ 12 $\mu$m that translates in $\Delta d \sim 60 $ nm giving an empirical error budget law of $\Delta d/\sigma_z\sim5 nm/\mu$m for systematic errors affecting the setup. Including the error budget from  $f$, the overall median uncertainty on the pinhole separation is $\Delta d \sim$ 75 nm (PV) for the direct fit of the fringes and $\Delta d \sim$ 110 nm (PV) for the PSD peak fit. The systematic errors noise floor prevents to probe the expected manufacturing precision range at 2 mm scale, $\sim$ 10 nm; nevertheless the 83 \% of the data points falls within the MICADO astrometric calibration requirement (dashed orange line) corresponding to an astrometric relative precision of 50 $\mu$as over 1 arcsec FoV.  
The manufacturing precision $\sigma_{pos}$ for the position of the pinholes on the WAM scales linearly with size $L$ of the mask [\citenum{Silvestri18}]:

\begin{equation}\label{imt_eq}
  \sigma_{pos} = \pm\left( 0.5\mu m + 5 \cdot 10^{-6} L\left[ m \right]\right)  
\end{equation}

Equation \ref{imt_eq} gives the absolute position error of the pinhole, while the relative position error of a pinhole with respect to an adjacent one is given by $\pm 5 \cdot 10^{-6} L\left[ m \right]$ that for a 1.5 mm pair gives $\sigma_{pos} = 7.5$ nm.

\section{Impact on MICADO astrometry calibration}
\label{cap_astrometry}

\begin{figure}
\plottwo{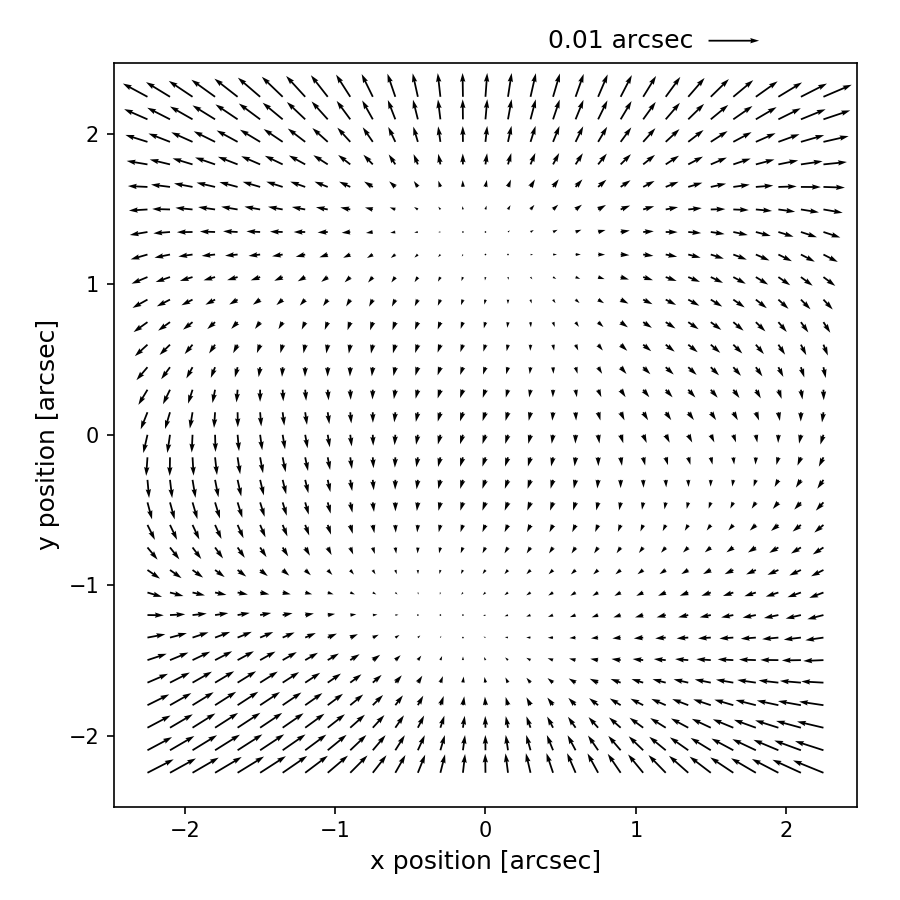}{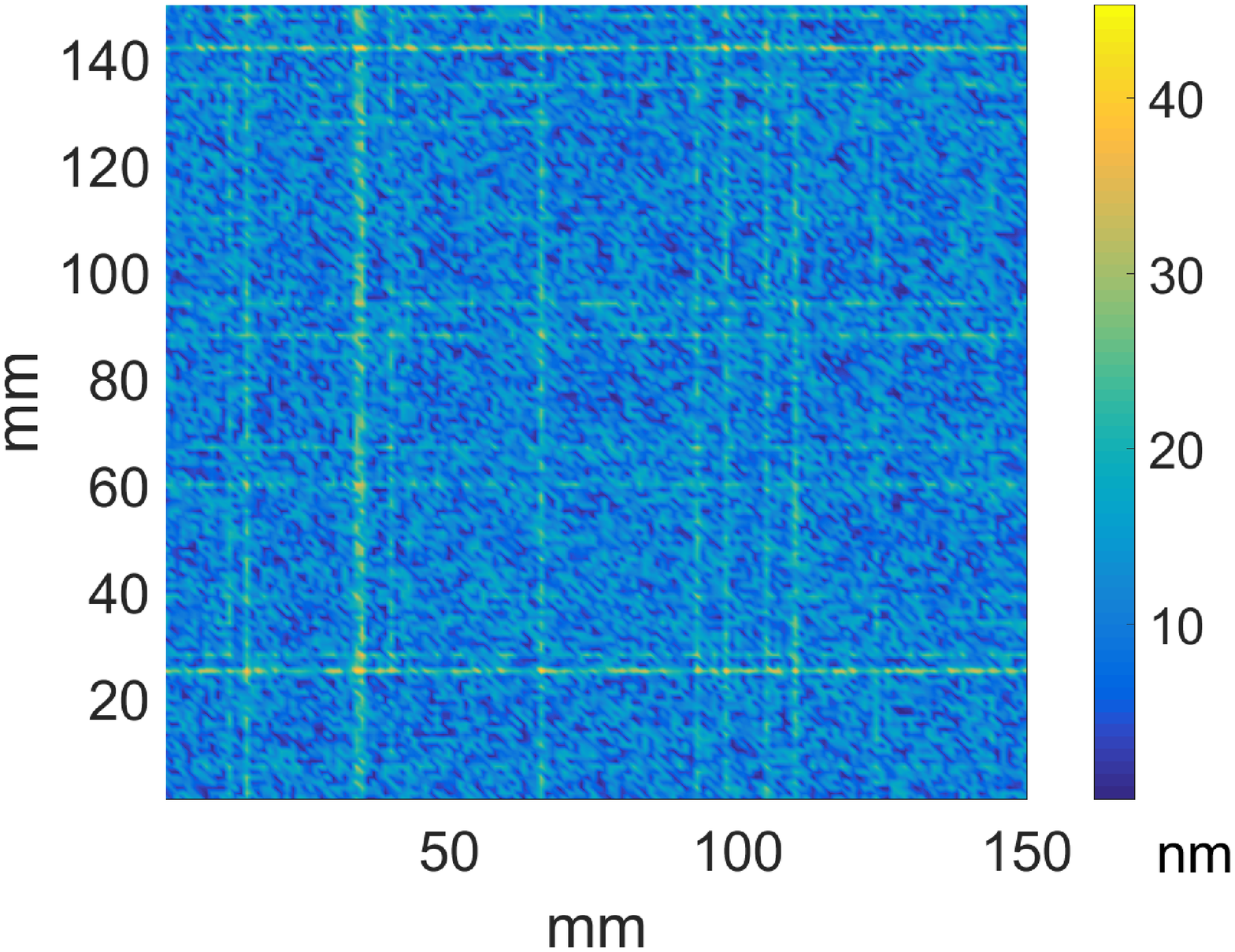}
\caption{Left: simulated MICADO distortion pattern in presence of MSF on the last mirror of the instrument close to the FP. The size of the field investigated is equivalent to $1/4^{th}$ of a HAWAII-4RG detector (the whole FP contains 9 detectors). Right: simulated MSF pattern from a power law 1/f$^2$ associated to the distortion pattern on the left, amplitude units in nanometer. \label{dist_map}}
\end{figure}

\begin{figure}
\epsscale{0.65}
\plotone{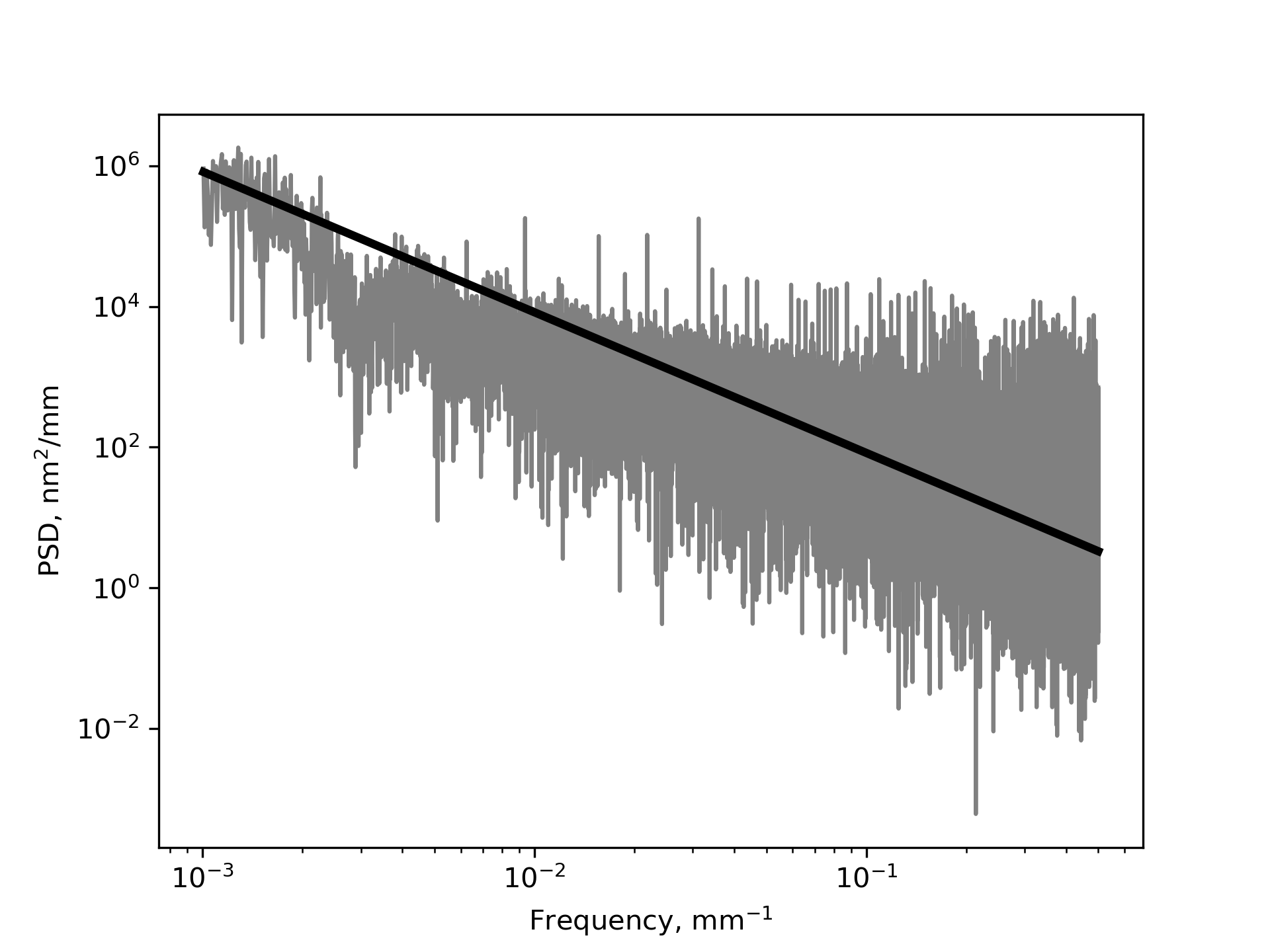}
\caption{Power spectrum distribution of the simulated MSF pattern shown in Fig. \ref{dist_map} and its linear fit with power law 1/f$^2$ (black line).\label{psd_tma}}
\end{figure}

The expected relative manufacturing precision in term of pinholes separation (Eq. \ref{imt_eq}) is $\sim$ 5nm/1mm. The test campaign on the WAM prototype, being limited by some setup systematic errors (see chapter \ref{cap_result}) has assessed the manufacturing precision down to $\sim$ 50nm/1mm scale. The pinhole position error increases with the linear range of manufacturing, remaining although very small ($\sim$ 2.5-5 nm) at the envisaged WAM pinholes pitch $\sim$ 0.5-1 mm. The WAM is densely populated of pinholes in order to fully sample the instruments distortion pattern down to the Mid Spatial Frequencies (MSF) scale. The latter depends on the manufacturing technology and grinding tool size; for the MICADO optics the expect MSF features (PV $\sim$ 50 nm) are expected to manifest at spatial scales of $\sim$ 5 mm being the approximate size of the manufacturing tool and to be characterized by a power law $1/f^2_{MSF}$ \citep{Sidick09} as shown in Fig. \ref{psd_tma}. In relation to MSF, all the optical surfaces close to the instrument intermediary or final Focal Plane (FP) are particularly important for the control of the distortions production. Close to the FPs in fact the optical beams from the different field points are converging towards the focus and their footprints on the surface depart one from each other leading to uncorrelated distortions between different fields. Since the MSF pattern is rapidly changing in space, the degree of the polynomial fit and the number of the required sampling points is large. This requirement, together with the need of a bright and dense point-like sources distribution justifies the use of astrometric masks for the calibration of the instruments. The manufacturing errors impact of the WAM at 10-100 nm scale on the calibration reliability has been verified using a series of ray tracing simulations in Zemax. We have reproduced a calibration scenario in presence of a series of MSF residuals on the last reflecting surface of MICADO close to the FP (Fig. \ref{dist_map}).

The distortion polynomial fit has been performed on a grid of input points randomized at 10-100 nm level versus the distorted coordinates of the same points at the instrument focal plane. Another fit has been run on a grid of nominal input points without any manufacturing error versus the same distorted coordinates associated to previous dataset where the manufacturing errors were taken into account. In this way, the latter case investigates the effect of not knowing or neglecting the manufacturing errors while assuming a perfect mask. The systematic post-fit astrometric RMS error that originates from neglecting these manufacturing errors is estimated at $\sim$ 2 $\mu$as (PV) as shown in Fig. \ref{dist_fourier}-left. The distortion pattern is efficiently fitted down to an RMS astrometric residual below the MICADO goal (50 $\mu$as) with a 10$^{th}$ order Legendre polynomial.

\textbf{\begin{figure}
\plottwo{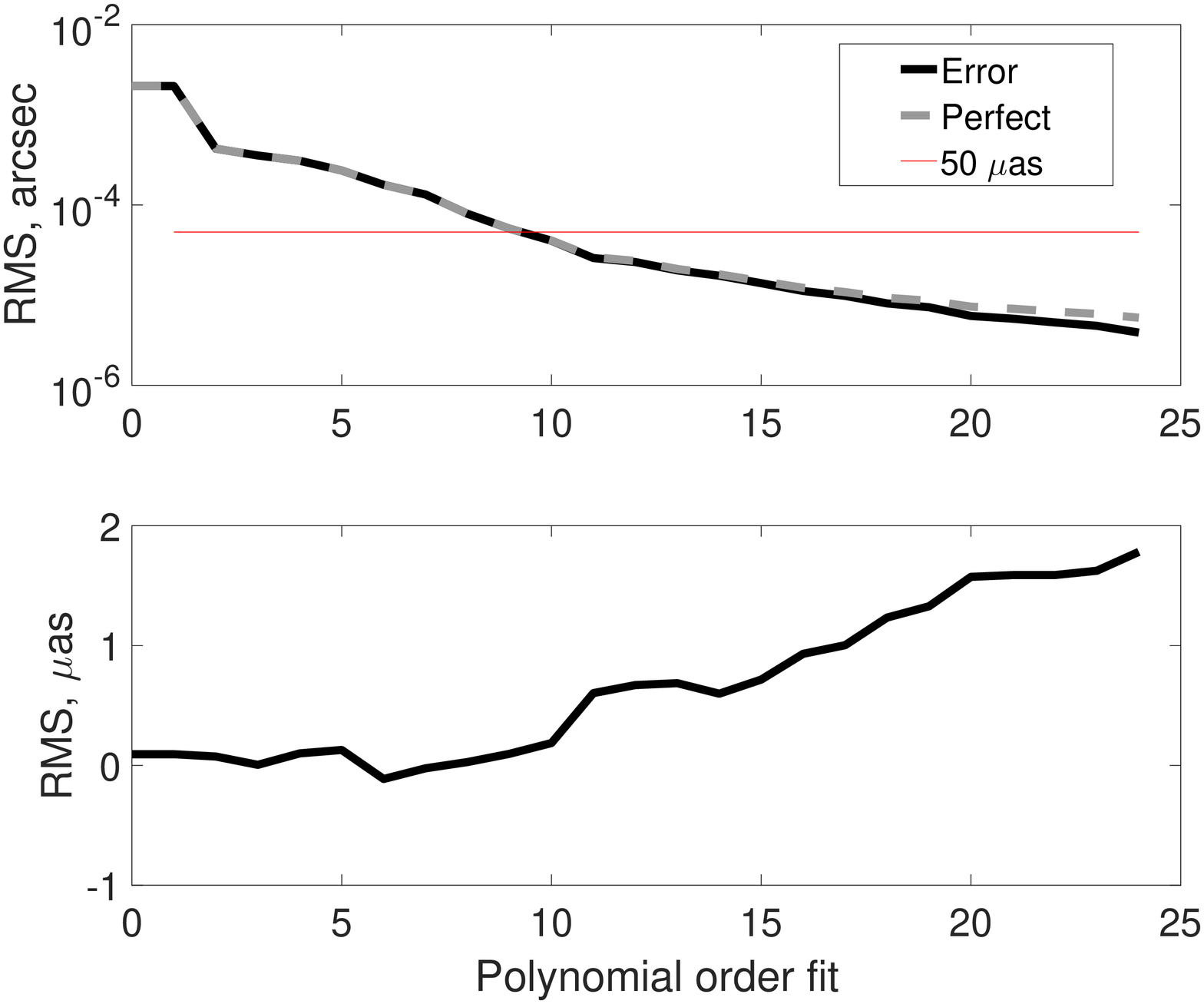}{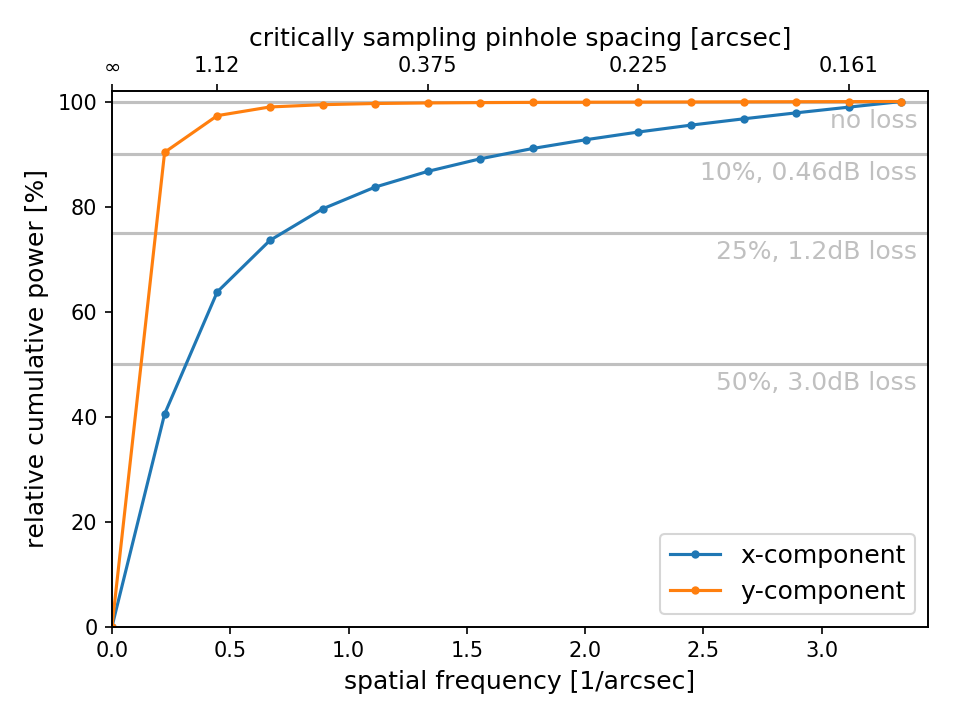}
\caption{Left-top: post-fit astrometric RMS residual error in presence of MSF on the mirror of the instrument close the FP; the red line indicates the 50 $\mu$as astrometric goal of MICADO. Left-bottom: systematic post-fit astrometric RMS error from neglecting the WAM manufacturing errors. Right: a minimum pinholes pitch of $\sim$ 0.5 mm is derived from the Fourier analysis of the critical sampling pinhole spacing. \label{dist_fourier}}
\end{figure}}

From the power spectrum analysis of the distortion pattern (Fig. \ref{dist_fourier}-right) we can infer the minimum pinholes pitch in order to sample completely the pattern in presence of MSF errors that based on the current analysis is $\sim$ 0.5 mm. This number is derived by multiplying the critical sampling pinhole spacing with the instrument plate scale. The distortion pattern can be additionally over-sampled by dithering the WAM at sub-mm scale using its hexapod fine positioning system.

\section{Conclusions}
\label{cap_conclusion}

The paper has described a revisited, modernized Young's experiment setup to measure at tens nanometer level the separation of pinholes pair on the prototype warm astrometric mask for MICADO. Although unable to achieve the ultimate manufacturing precision of the mask $\sim$ 5nm/1mm, the current setup has proved the reliability of the lithographic technology within the MICADO astrometric requirement. Having assessed the reliability of the experimental setup down to $\sim$ 50nm/1mm scale, it can in be used for future characterizations of different mask prototypes and manufacturing technologies in the range of interest for the calibration of the MICADO astrometry. The measurement has led to confirm the manufacturing reliability of the WAM in relation to the MICADO astrometric requirement $\sim$ 50nm/1mm. 
We studied at a simulation level the MICADO distortion pattern in presence of MSF and of 10-100 nm manufacturing errors on the WAM deriving a minimum pinholes pitch of $\sim$ 0.5 mm to fully sample the spatial features of the pattern. We demonstrated that the MICADO astrometric precision of 50 $\mu$as is achievable also in presence of a MSF pattern and manufacturing errors of the WAM with the use of a 10$^{th}$ order Legendre polynomial for distortion fit.

\section{Acknowledgments}

Authors are grateful to Luciano Silvestri and Ren\'e Thoma from IMT for the  support in developing the WAM prototype, and to Florian Kerber and Jeroen Bouwman for the support in the study of the coating aging of the mask and pattern recognition respectively.



\end{document}